\documentstyle[amsmath,amssymb,graphicx,wrapfig]{article}

\def\be{\begin{eqnarray}}
\def\ee{\end{eqnarray}}

\hoffset= - 1.0in         

\begin{document}

\baselineskip14pt

\hfill ITEP/TH-78/09

\bigskip

\begin{center}
{\LARGE Exact solution for mean energy of 2d Dyson gas at $\beta = 1$ }\\
\vspace{0.7cm}
{\large Sh.Shakirov }\\
\vspace{0.3cm}
{\em ITEP, Moscow, Russia\\
MIPT, Dolgoprudny, Russia}\\
\end{center}

\bigskip

Mean Coulomb energy of 2d Dyson gas in quadratic potential is examined from combinatorial viewpoint. For $\beta = 1$, we find a recursive relation on mean energy and obtain its exact (finite $N$) solution in closed form in terms of the hypergeometric function $_{3}F_{2}$. Using this exact solution, we derive the large-$N$ asymptotic expansion of mean energy and show, that this expansion contains half-integer powers of $N$.

\section{Introduction}

This paper is devoted to the study of a system of $N$ charged particles on the complex plane with complex coordinates $z_i$, which are bound to the origin by attractive Gaussian potential and interact with each other by pairwise repulsive Coulomb interactions. In the low energy approximation (neglecting the kinetic energy of particles) partition function of such a system can be written as a $2N$-fold integral

\begin{align}
Z_{N}(\beta) = \dfrac{1}{\pi^N} \int d^2z_1 \ldots d^2z_N \exp\left( - \sum\limits_{i} \big|z_i\big|^2 + \beta \sum\limits_{i < j} \log \left| z_i - z_j \right|^2 \right)
\label{PartitionFunction}
\end{align}
\smallskip\\
where $\beta$ determines the strength of Coulomb repulsion. It is conventional to rewrite $Z_{N}(\beta)$ as

\begin{align}
Z_{N}(\beta) = \dfrac{1}{\pi^N} \int d^2z_1 \ldots d^2z_N \prod\limits_{i < j} \left| z_i - z_j \right|^{2\beta} \exp\Big( - \sum\limits_{i} \big|z_i\big|^2 \Big)
\end{align}
\smallskip\\
and to define the correlators (statistical averages in the model) via

\begin{align}
\Big< f( z_1, \ldots, z_N ) \Big> = \dfrac{1}{\pi^N Z_{N}(\beta)} \int d^2z_1 \ldots d^2z_N \ f(z_1, \ldots, z_N ) \ \prod\limits_{i < j} \left| z_i - z_j \right|^{2\beta} \exp\Big( - \sum\limits_{i} \big|z_i\big|^2 \Big)
\label{Correlators}
\end{align}
\smallskip\\
This physical system, known as 2d Dyson gas \cite{DysonGas}, is quite interesting for a number of reasons. First of all, it is expected to undergo a phase transition at certain large value of the Coulomb charge $\beta$ \cite{IZ, PhaseTransition}. Second, its thermodynamic limit (i.e, the limit of large $N$) is known to be related to growth processes, such as Laplacian growth \cite{LaplacianGrowth}. The last but not the least, 2d Dyson gas is a typical representative example of $\beta$-ensembles \cite{BetaEnsembles}, which today attract increasing attention because of their recently established relation \cite{RelAGT} to conformal field theory and the AGT conjecture \cite{AGT}. In this context $\beta$-ensembles are used to describe conformal blocks of Virasoro and W-algebras and equivalent Nekrasov functions. The parameter $\beta$ is directly related to the central charge $c$ in conformal field theory and to Nekrasov parameters $\epsilon_1, \epsilon_2$ in supersymmetric gauge theory. All this indicates, that dependence of $Z_N$ on $\beta$ is non-trivial, interesting and worth investigating.

\pagebreak

One of the simplest ways to study this dependence is to consider $Z_{N}(\beta)$ as a Taylor series near $\beta = 1$

\begin{align}
\log Z_N(\beta) = \log Z_N(1) + (\beta - 1) E_N(1) + \dfrac{(\beta - 1)^2}{2!} \ H_N(1) + \ldots
\label{Expansion}
\end{align}
\smallskip\\
and calculate several first coefficients in this expansion as correlators

\begin{align}
E_N(1) = \left< \sum\limits_{i < j} \log \left| z_i - z_j \right|^2 \right>
\end{align}
\begin{align}
H_N(1) = \left< \sum\limits_{i < j} \sum\limits_{k < l} \log \left| z_i - z_j \right|^2 \log \left| z_k - z_l \right|^2 \right> - \left< \sum\limits_{i < j} \log \left| z_i - z_j \right|^2 \right> \left< \sum\limits_{k < l} \log \left| z_k - z_l \right|^2 \right>
\end{align}
\smallskip\\
evaluated at the point $\beta = 1$. The point $\beta = 1$ is distinquished, because precisely at this point the system is equivalent to the matrix model of complex matrices \cite{Complex}. Correlators $E_N$ and $H_N$ have a physical meaning of average Coulomb energy and heat, respectively. In this paper, we find  the simplest of these coefficients $E_N$ by combinatorial methods. We find that $E_N$ satisfies certain closed recursive relation with respect to $N$, solve this relation explicitly and obtain an exact formula for $E_N$. We then derive its large-$N$ asymptotics, which may be of interest in particular physical applications.

\section{Partition function}

Let us note, that a wide class of correlators in 2d model can be explicitly evaluated using a simple rule

\begin{align}
\int d^2 z \ z^i {\bar z}^j e^{- z{\bar z}} = \pi \ \delta_{ij} \ \Gamma(1 + i)
\label{Rule}
\end{align}
\smallskip\\
One of the simplest examples of such evaluations is the partition function itself at $\beta = 1$:

\begin{align}
Z_N(1) = \dfrac{1}{\pi^N} \int d^2z_1 \ldots d^2z_N \prod\limits_{i < j} ( z_i - z_j ) \prod\limits_{i < j} ( {\bar z}_i - {\bar z}_j ) \exp\Big( - \sum\limits_{i} \big|z_i\big|^2 \Big)
\end{align}
\smallskip\\
Opening the brackets in the Vandermonde determinant and using the rule (\ref{Rule}), it is easy to find

\begin{equation}
\addtolength{\fboxsep}{5pt}
\boxed{
\begin{gathered}
Z_N(1) = \prod\limits_{k = 0}^{N} \Gamma(1 + k), \ \ \ \ \ \log Z_N(1) = \sum\limits_{k = 1}^{N+1} \log \Gamma(1 + k)
\end{gathered}
}\label{PartitionFunctionFinal}
\end{equation}
\smallskip\\
From the point of view of the Taylor expansion (\ref{Expansion}), this provides an exact formula for the 0th term. The rest of this paper is devoted to a similar (though technically more involved) calculation of the 1st term of the expansion (\ref{Expansion}). Higher order terms would be also very interesting to compute; however, they are more complicated and remain beyond the scope of present paper. They will be considered elsewhere.

\section{Mean energy}

We find it convenient to evaluate the mean energy

\begin{align}
E_N(1) = \dfrac{1}{ \pi^N Z_{N}\big(1\big) } \int d^2z_1 \ldots d^2z_N \left( \sum\limits_{i < j} \log \left| z_i - z_j \right|^{2} \right) \ \prod\limits_{i < j} \left| z_i - z_j \right|^{2} \exp\Big( - \sum\limits_{i} \big|z_i\big|^2 \Big)
\label{EnergyDefinition}
\end{align}
\smallskip\\
using the replica limit

\begin{align}
E_N(1) = \lim\limits_{\epsilon \rightarrow 0} \dfrac{e_{N}(\epsilon) - 1}{\epsilon}, \ \ \ \ \ \ e_N(\epsilon) = \left< \sum\limits_{i < j} \left| z_i - z_j \right|^{2 \epsilon} \right>
\end{align}
\smallskip\\
where the correlator

\begin{align}
e_{N}(\epsilon) = \dfrac{1}{  \pi^N Z_{N}\big(1\big) } \int d^2z_1 \ldots d^2z_N \left( \sum\limits_{i < j} \left| z_i - z_j \right|^{2\epsilon} \right) \ \prod\limits_{i < j} \left| z_i - z_j \right|^{2} \exp\Big( - \sum\limits_{i} \big|z_i\big|^2 \Big)
\end{align}
\smallskip\\
can be easily calculated for natural $\epsilon$ with the help of (\ref{Rule}). We present this calculation here in some detail. Since the integrand is symmetric under all permutations of $z_1, \ldots, z_N$, we have

\begin{align}
e_{N}(\epsilon) = \dfrac{N(N-1)}{2 \pi^N Z_N(1)} \int d^2z_1 \ldots d^2z_N \left| z_1 - z_2 \right|^{2\epsilon} \prod\limits_{i < j} \left| z_i - z_j \right|^{2} \exp\Big( - \sum\limits_{i} \big|z_i\big|^2 \Big)
\end{align}
\smallskip\\
Expanding the Vandermonde determinant, we find

\begin{align}
\prod\limits_{i < j} \left| z_i - z_j \right|^{2} = {\cal E}_{i_1 \ldots i_N} {\cal E}_{j_1 \ldots j_N} z_1^{i_1} \ldots z_N^{i_N} {\bar z}_1^{j_1} \ldots {\bar z}_N^{j_N}
\label{Vandermonde}
\end{align}
\smallskip\\
where the two sums over $(i_1, \ldots, i_N)$ and $(j_1, \ldots, j_N)$ are not written explicitly but assumed. Both sums are taken over permutations of $(0,\ldots, N-1)$, and ${\cal E}$ is the sign of permutation. Using (\ref{Vandermonde}) we obtain

\begin{align}
e_{N}(\epsilon) =  \dfrac{N(N-1)}{2 \pi^N Z_N(1)} \ {\cal E}_{i_1 \ldots i_N} {\cal E}_{j_1 \ldots j_N} \int d^2z_1 \ldots d^2z_N \left| z_1 - z_2 \right|^{2\epsilon} z_1^{i_1} \ldots z_N^{i_N} {\bar z}_1^{j_1} \ldots {\bar z}_N^{j_N} \exp\Big( - \sum\limits_{i} \big|z_i\big|^2 \Big)
\end{align}
\smallskip\\
Integrating out variables $z_3, \ldots, z_N$ with the help of integration rule (\ref{Rule}) we find

\begin{align}
e_{N}(\epsilon) = \dfrac{N(N-1)}{2 \pi^2 Z_N(1)} \ {\cal E}_{i_1 \ldots i_N} {\cal E}_{j_1 \ldots j_N}  \prod\limits_{k = 3}^{N} i_k! \delta^{i_k}_{j_k} \ \int d^2z_1 d^2z_2 \left| z_1 - z_2 \right|^{2\epsilon} z_1^{i_1} z_2^{i_2} {\bar z}_1^{j_1} {\bar z}_2^{j_2} \exp\Big( - \big|z_1\big|^2 - \big|z_2\big|^2 \Big)
\label{Prev}
\end{align}
\smallskip\\
With the help of the same rule (\ref{Rule}), the partition function $Z_N(1)$ can be calculated:

\begin{align}
\nonumber Z_N(1) \ = \ & \dfrac{1}{\pi^N} \int d^2z_1 \ldots d^2z_N \prod\limits_{i < j} \left| z_i - z_j \right|^{2\beta} \exp\Big( - \sum\limits_{i} \big|z_i\big|^2 \Big) = \\ \nonumber & \\ \nonumber &
= \dfrac{1}{\pi^N} {\cal E}_{i_1 \ldots i_N} {\cal E}_{j_1 \ldots j_N} \int d^2z_1 \ldots d^2z_N z_1^{i_1} \ldots z_N^{i_N} {\bar z}_1^{j_1} \ldots {\bar z}_N^{j_N} \exp\Big( - \sum\limits_{i} \big|z_i\big|^2 \Big) = \prod\limits_{k = 0}^{N} k!
\label{PartFunc}
\end{align}
\smallskip\\
Substituting (\ref{PartFunc}) into (\ref{Prev}), we obtain

\begin{align}
e_{N}(\epsilon) = \dfrac{N(N-1)}{2 \pi^2} \dfrac{{\cal E}_{i_1 i_2 i_3 \ldots i_N} {\cal E}_{j_1 j_2 i_3 \ldots i_N}  i_3! \ldots i_N!}{0! 1! \ldots N!} \int d^2z_1 d^2z_2 \left| z_1 - z_2 \right|^{2\epsilon} z_1^{i_1} z_2^{i_2} {\bar z}_1^{j_1} {\bar z}_2^{j_2} \exp\Big( - \big|z_1\big|^2 - \big|z_2\big|^2 \Big)
\end{align}
\smallskip\\
After minor simplification, this expression takes form

\begin{align}
e_{N}(\epsilon) = \dfrac{1}{\pi^2} \sum\limits_{0 \leq a < b < N} \dfrac{1}{a!b!} \int d^2z_1 d^2z_2 \left| z_1 - z_2 \right|^{2\epsilon} \Big( z_1^{a} z_2^{b} {\bar z}_1^{a} {\bar z}_2^{b} - z_1^{a} z_2^{b} {\bar z}_1^{b} {\bar z}_2^{a} \Big) \exp\Big( - \big|z_1\big|^2 - \big|z_2\big|^2 \Big)
\end{align}
\smallskip\\
At natural $\epsilon$, this correlator can be calculated by opening the brackets

\begin{align}
\left| z_1 - z_2 \right|^{2\epsilon} = \sum\limits_{i,j = 0}^{\epsilon} \dfrac{\epsilon!}{i!(\epsilon - i)!} \dfrac{\epsilon!}{j!(\epsilon - j)!}  (-1)^{i+j} z_1^{\epsilon - i} z_2^{i} {\bar z}_1^{\epsilon - j} {\bar z}_2^{j}
\end{align}
\smallskip\\
and using the same integration rule (\ref{Rule}). In this way we find

\begin{figure}[t]
\begin{center}
$
\begin{array}{c|cccccc}
e_N(\epsilon) & \epsilon = 1 & \epsilon = 2 & \epsilon = 3 & \epsilon = 4 & \epsilon = 5 \\ \\
\hline\\
N = 2 & 4& 24& 192& 1920& 23040\\ \\
N = 3 & 15& 114& 1152& 14400& 213120\\ \\
N = 4 & 36& 332& 4056& 60720& 1064640\\ \\
N = 5 & 70& 760& 10890& 189720& 3838200\\ \\
N = 6 & 120& 1500& 24660& 489744& 11218320\\ \\
N = 7 & 189& 2674& 49602& 1105608& 28268520\\ \\
\end{array}
$
\caption{Several first values of $e_N(\epsilon)$, calculated with the help of (\ref{Proof}). }
\end{center}
\end{figure}
\begin{align}
e_{N}(\epsilon) = \sum\limits_{0 \leq a < b < N} \left( \sum\limits_{i = 0}^{\epsilon} \dfrac{\epsilon!^2}{i!^2(\epsilon - i)!^2} - \sum\limits_{i = 0}^{\epsilon-b+a} \dfrac{(-1)^{a-b} \epsilon!}{i!(\epsilon - i)!} \dfrac{\epsilon!}{(i + b - a)!(\epsilon - i - b + a)!} \right) \dfrac{(\epsilon - i + a)! (i + b)!}{a!b!}
\label{Proof}
\end{align}
\smallskip\\
Note, that for each particular natural $N$ and $\epsilon$ formula (\ref{Proof}) is just a finite sum, which can be evaluated exactly. Given this combinatorial formula, one straightforwardly proves that $e_{N}(\epsilon)$ satisfies a recursion

\begin{align}
\nonumber & (N+1)(\epsilon+2 N+3)(\epsilon+2 N+4) e_{N}(\epsilon) - \emph{} \\ \nonumber & \\ \nonumber  & \emph{} - 2 (12+34 N+26 N^2+6 N^3+5 \epsilon+10 N \epsilon+4 N^2 \epsilon+\epsilon^2+N \epsilon^2) e_{N+1}(\epsilon) + \emph{} \\ \nonumber  & \\ \nonumber  & \emph{} + (N+1) (12 N^2+38 N+4 N \epsilon+\epsilon^2+5 \epsilon+24) e_{N+2}(\epsilon) - \emph{} \\ \nonumber  & \\ & \emph{} - 4 (N+2) (N+1)^2 e_{N+3}(\epsilon) = 0
\end{align}
\smallskip\\
The proof is just by substitution. This is very useful: expanding now $e_{N}(\epsilon)$ in powers of $\epsilon$

\begin{align}
e_{N}(\epsilon) = \dfrac{N(N-1)}{2} + \epsilon E_N(1) + \ldots
\end{align}
\smallskip\\
we find a recursion, satisfied by the mean Coulomb energy:

\begin{align}
\nonumber & 2 (2 N+3) (N+2) (N+1) E_N(1) - \emph{} \\ \nonumber & \\ \nonumber & \emph{} - 4 (N+2) (3 N^2+7 N+3) E_{N+1}(1) + \emph{} \\ \nonumber & \\ \nonumber & \emph{} + 2 (N+1) (6 N^2+19 N+12) E_{N+2}(1) - \emph{} \\ \nonumber & \\ & \emph{} - 4 (N+2) (N+1)^2 E_{N+3}(1) + (3 N+5) (N+1) = 0
\end{align}
\smallskip\\
The general 3-parametric solution of this linear 3-rd order difference equation is

\begin{align}
E_{N}(1) = \dfrac{1}{2} N^2 \Psi(N) + \dfrac{1}{4} + c_1 N + c_2 N^2 + c_3 \sum\limits_{k = 2}^{N-1} \dfrac{N(N-1)}{k(k-1)} \dfrac{1}{(k+1)!} \dfrac{\Gamma\left(k + \dfrac{3}{2}\right)}{\Gamma\left(\dfrac{3}{2}\right)}
\label{GenericSolutionEnergy}
\end{align}
\smallskip\\
where $\Psi(N) = \sum_{k=1}^{N} 1/k$. Eq. (\ref{GenericSolutionEnergy}) can be easily proven by direct substitution. From the initial conditions

\[
\left\{
\begin{array}{lll}
E_{2}(1) = 1 - \gamma + \ln 2 \\
\\
E_{3}(1) = \dfrac{29}{8} - 3 \gamma + 3 \ln 2 \\
\\
E_{4}(1) = \dfrac{793}{96} - 6 \gamma + 6 \ln 2\\
\\
\end{array}
\right.
\]
we find the following values of the constants $c_1,c_2,c_3$:

\[
\left\{
\begin{array}{lll}
c_1 = \dfrac{13}{8} + \dfrac{1}{2} \gamma - \dfrac{1}{2} \ln 2  \\
\\
c_2 = - \dfrac{9}{8} + \dfrac{1}{2} \ln 2 \\
\\
c_3 = 1 \\
\end{array}
\right.
\]
\smallskip\\
Therefore, we obtain an exact formula for mean Coulomb energy at finite $N$:

\begin{align*}
E_{N}(1) \ = \ & \dfrac{1}{2} N^2 \Psi(N) + \dfrac{1}{4} + \left( \dfrac{13}{8} + \dfrac{1}{2} \gamma - \dfrac{1}{2} \ln 2 \right) N + \left( - \dfrac{9}{8} + \dfrac{1}{2} \ln 2 \right) N^2 + \emph{} \\ & \\ & \emph{} + \sum\limits_{k = 2}^{N-1} \dfrac{N(N-1)}{k(k-1)} \dfrac{1}{(k+1)!} \dfrac{\Gamma\left(k + \dfrac{3}{2}\right)}{\Gamma\left(\dfrac{3}{2}\right)}
\end{align*}
\smallskip\\
The remaining sum can be evaluated in terms of the hypergeometric function $_{3}F_{2}$:

\begin{align*}
\sum\limits_{k = 2}^{N-1} \dfrac{N(N-1)}{k(k-1)} \dfrac{1}{(k+1)!} \dfrac{\Gamma\left(k + \dfrac{3}{2}\right)}{\Gamma\left(\dfrac{3}{2}\right)} = \dfrac{N(N-1)}{2} \left( \dfrac{7}{4} - \ln 2 \right) - \dfrac{ \Gamma\left( N + \dfrac{3}{2} \right)}{\Gamma\left( N + 2 \right) \Gamma\left( \dfrac{3}{2} \right)} {}_{3}F_{2} \left( \left. \begin{array}{ccc} 1, N - 1, N + 3/2 \\ N + 2, N + 1 \end{array} \right| \ 1 \ \right)
\end{align*}
\smallskip\\
Using this formula, we finally obtain a relatively simple answer:

\begin{equation}
\addtolength{\fboxsep}{5pt}
\boxed{
\begin{gathered}
E_{N}(1) = \dfrac{N^2 \Psi(N)}{2} - \dfrac{N^2}{4} + \dfrac{3N}{4} + \dfrac{1}{4} + \dfrac{N \gamma}{2} - \dfrac{ \Gamma\left( N + 3/2 \right)}{\Gamma\left( N + 2 \right) \Gamma\left( 3/2 \right)} \ {}_{3}F_{2} \left( \left. \begin{array}{ccc} 1, N - 1, N + 3/2 \\ N + 2, N + 1 \end{array} \right| \ 1 \ \right)
\end{gathered}
}\label{AverageEnergyFinal}
\end{equation}
\smallskip\\
If one rescales partition function

\begin{align}
{\tilde Z}_{N}(\beta) = \dfrac{1}{\pi^N} \int d^2z_1 \ldots d^2z_N \prod\limits_{i < j} \left| z_i - z_j \right|^{2\beta} \exp\Big( - N \sum\limits_{i} \big|z_i\big|^2 \Big) = \left( \dfrac{1}{\sqrt{N}} \right)^{\beta N(N-1) + 2N} Z_{N}(\beta)
\end{align}
\smallskip\\
then one also needs to shift the energy:

\begin{align}
{\tilde E}_{N}(1) = E_{N}(1) - N(N-1)/2 \ln N
\label{AverageEnergyFinal2}
\end{align}
\smallskip\\
\begin{figure}\begin{center}
\includegraphics[totalheight=200pt]{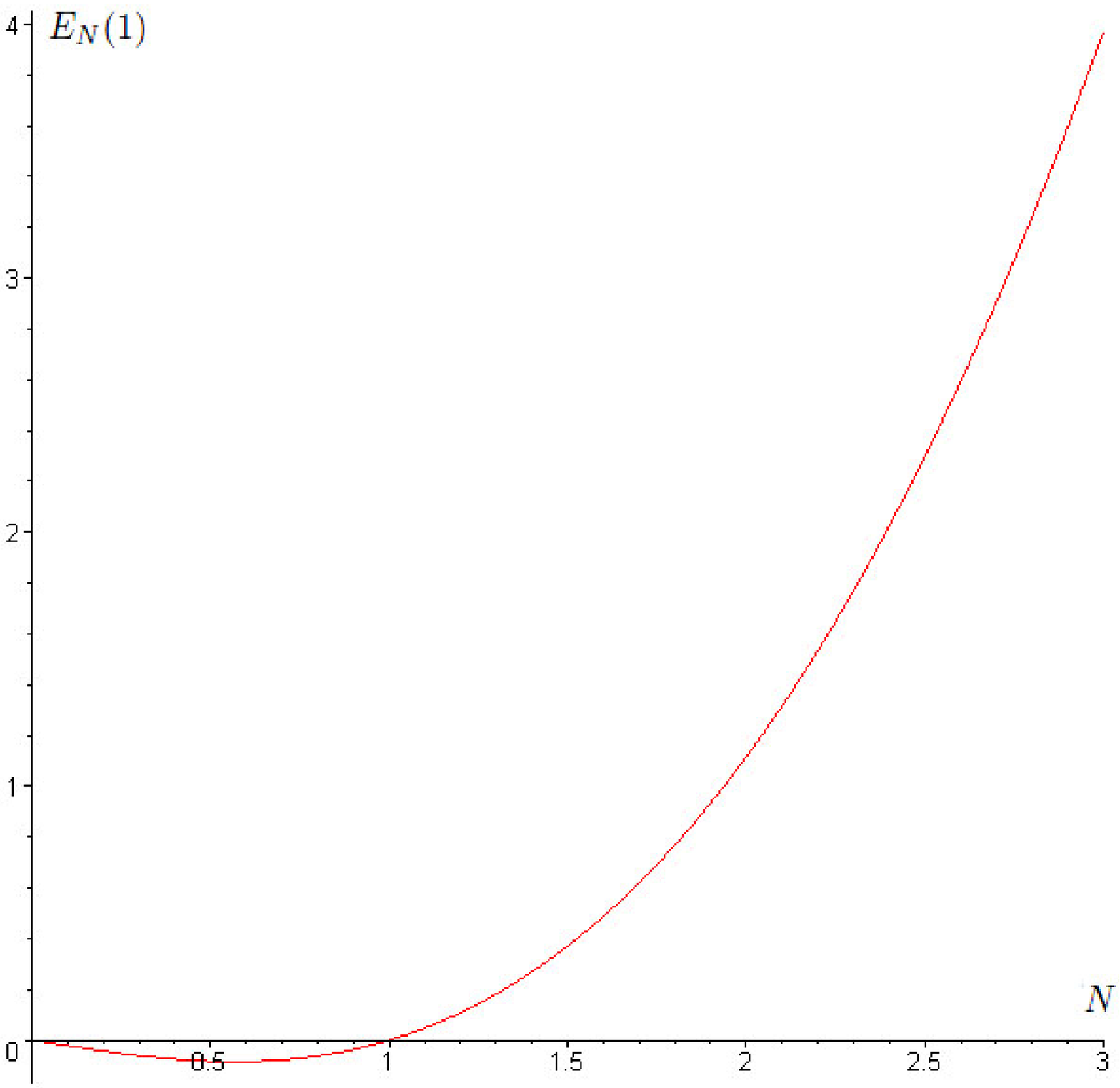}
\caption{\footnotesize{
Mean Coulomb energy (\ref{AverageEnergyFinal}) for small $N$.
}}\label{EnergyGraph}
\end{center}\end{figure}
The values of ${\tilde E}_{N}(1)$, calculated with this formula, coincide with Monte-Carlo results, obtained in \cite{MonteCarlo}:

\[
\begin{array}{c|c|c|c|c|c}
N & \mbox{Exact answer via } (\ref{AverageEnergyFinal2}) & \mbox{Its evaluation to 20 digits} & \mbox{Monte-Carlo result} & \mbox{Monte-Carlo error} \\
& & & & & \\
\hline& & & & & \\
5 & \dfrac{5831}{384} - 10 \gamma + 10 \ln\left(\dfrac{2}{5}\right) & 0.249831865576454075 & 0.25013 & 0.00030 \\
\hline& & & & & \\
10 & (\ldots) - 45 \gamma + 45 \ln\left(\dfrac{2}{10}\right) & -7.766078379434240251 & -7.76667 & 0.00065 \\
\hline& & & & & \\
25 & (\ldots) - 300 \gamma + 300 \ln\left(\dfrac{2}{25}\right)  & -99.84816319738693935 & -99.85156 & 0.00329 \\
\hline& & & & & \\
50 & (\ldots) - 1225 \gamma + 1225 \ln\left(\dfrac{2}{50}\right)  & -492.8773199345421588 & -492.88918  & 0.00702  \\
\hline& & & & & \\
75 & (\ldots) - 2775 \gamma + 2775 \ln\left(\dfrac{2}{75}\right)  & -1191.5028305194359964 & -1191.49356 & 0.00974 \\
\hline& & & & & \\
\end{array}
\]
where $(\ldots)$ stand for certain rational numbers, which we do not write explicitly due to space limitations. This good agreement with numeric calculations supports validity of the main formula (\ref{AverageEnergyFinal}).

\section{Large $N$ expansion of the mean energy}

Let us now obtain the large-$N$ expansion of the mean energy (\ref{AverageEnergyFinal}). Since

\begin{align*}
\dfrac{1}{k(k-1)} \dfrac{1}{(k+1)!} \dfrac{\Gamma\left(k + \dfrac{3}{2}\right)}{\Gamma\left(\dfrac{3}{2}\right)} = \dfrac{2}{\sqrt{\pi}} k^{-5/2} + \dfrac{3}{4\sqrt{\pi}} k^{-7/2} + \dfrac{121}{64\sqrt{\pi}} k^{-9/2} + \ldots
\end{align*}
\smallskip\\
we have

\begin{align*}
\sum\limits_{k = 2}^{N-1} \dfrac{1}{k(k-1)} \dfrac{1}{(k+1)!} \dfrac{\Gamma\left(k + \dfrac{3}{2}\right)}{\Gamma\left(\dfrac{3}{2}\right)} = {\rm const} - \dfrac{4}{3\sqrt{\pi}} N^{-3/2} + \dfrac{7}{10\sqrt{\pi}} N^{-5/2} - \dfrac{391}{672\sqrt{\pi}} N^{-7/2} + \ldots
\end{align*}
\smallskip\\
where the constant is easy to find:

\begin{align*}
{\rm const} = \sum\limits_{k = 2}^{\infty} \dfrac{1}{k(k-1)} \dfrac{1}{(k+1)!} \dfrac{\Gamma\left(k + \dfrac{3}{2}\right)}{\Gamma\left(\dfrac{3}{2}\right)} = \dfrac{7}{8} - \dfrac{1}{2} \ln 2
\end{align*}
\smallskip\\
Consequently,

\begin{align*}
\sum\limits_{k = 2}^{N-1} \dfrac{N(N-1)}{k(k-1)} \dfrac{1}{(k+1)!} \dfrac{\Gamma\left(k + \dfrac{3}{2}\right)}{\Gamma\left(\dfrac{3}{2}\right)} = N(N-1) \left( \dfrac{7}{8} - \dfrac{1}{2} \ln 2 \right) - \dfrac{4}{3\sqrt{\pi}} N^{1/2} + \dfrac{1}{30\sqrt{\pi}} N^{-1/2} - \dfrac{107}{3360\sqrt{\pi}} N^{-3/2} + \ldots
\end{align*}
\smallskip\\
Similarly

\begin{align}
\dfrac{N^2}{2} \Psi(N) = \dfrac{1}{2} N^2 \ln N - \dfrac{N}{4} - \dfrac{1}{24} + \dfrac{1}{240} N^{-2} - \dfrac{1}{504} N^{-4} + \ldots
\end{align}
\smallskip\\
After all cancelations and simplifications, we find the large-$N$ expansion of (\ref{AverageEnergyFinal})

\begin{equation}
\addtolength{\fboxsep}{5pt}
\boxed{
\begin{gathered}
E_{N}(1) = \dfrac{1}{2} N^2 \ln N - \dfrac{N^2}{4} + \left( \dfrac{1}{2} + \dfrac{1}{2} \gamma \right) N + \dfrac{5}{24} - \dfrac{4  N^{1/2}}{3\sqrt{\pi}} + \dfrac{N^{-1/2}}{30\sqrt{\pi}}  - \dfrac{107 N^{-3/2}}{3360\sqrt{\pi}} + \dfrac{N^{-2}}{240}  + \ldots
\end{gathered}
}\label{AsymptoticsFinal}
\end{equation}
\smallskip\\
which contains, except for the leading logarithmic asymptotics, integer and half-integer powers of $N$.

\section{Conclusion}

In this paper, we have obtained an exact finite $N$ formula (\ref{AverageEnergyFinal}) for the mean energy of $\beta = 1$ 2d Dyson gas and its large $N$ expansion (\ref{AsymptoticsFinal}). This mean energy describes the 2d Dyson gas partition function $Z_N(\beta)$ in the first order of the $(\beta - 1)$ expansion. It would be interesting to extend these results to higher orders of the $(\beta - 1)$ expansion. In this direction, the next simplest quantity after the mean energy is the mean heat $H_{N}(1)$. In analogue with $E_{N}(1)$, we expect $H_{N}(1)$ to satisfy some recursive relation in $N$. This relation, as well as its exact solution, remains to be found and investigated. It is also desirable to find a conceptual explanation for appearance of such recursive relations. Probably, the origin of these recursive relations lies in the integrable structure \cite{Integrability} of the complex matrix model.

\section{Acknowledgements}

We are indebted to A.Zabrodin for illuminating discussions. Our work is partly supported by Russian Federal Nuclear Energy Agency, Russian Federal Agency for Science and Innovations under contract 02.740.11.5029 and the Russian President's Grant of Support for the Scientific
Schools NSh-3035.2008.2, by RFBR grant 07-02-00645, by the joint grants 09-02-90493-Ukr, 09-01-92440-CE,
09-02-91005-ANF and 09-02-93105-CNRS. The work of Sh.Shakirov is also supported in part by the Moebius
Contest Foundation for Young Scientists.

\end{document}